\documentclass[aps,prd,reprint,superscriptaddress]{revtex4-2}
\usepackage{slashed}
\usepackage{amsmath}
\usepackage{graphicx}
\usepackage{mathtools}
\usepackage[caption=false,position=top]{subfig}
\usepackage[export]{adjustbox}
\usepackage{dsfont}
\usepackage{xcolor}
\usepackage{lineno}
\usepackage{soul}
\usepackage{hyperref}
\hypersetup{colorlinks=true,linkcolor=black,citecolor=blue,urlcolor=blue}

\begin{document}

\title{Gravitationally localized states of two neutral fermions interacting with a Higgs field}


\author{Peter E.~D.~Leith}
\author{Alasdair Dorkenoo Leggat}
\author{Chris A.~Hooley}
\author{Keith Horne}
\affiliation{SUPA, School of Physics and Astronomy, University of St Andrews, North Haugh, St Andrews, Fife KY16 9SS, UK}
\author{David G.~Dritschel}
\affiliation{School of Mathematics and Statistics, University of St Andrews, North Haugh, St Andrews, Fife KY16 9SS, UK}


\date{\today}

\begin{abstract}
We present localized `particle-like' states composed of a pair of neutral fermions interacting with a scalar Higgs field and the metric of spacetime, extending the Einstein--Dirac formalism introduced by Finster, Smoller, and Yau [Phys.~Rev.~D \textbf{59}, 104020 (1999)]. We demonstrate that, when the coupling between the fermions and the Higgs field is strong, there is a class of states in which the total (ADM) mass no longer increases proportionally to the mass of the constituent fermions; indeed it decreases. This phenomenon enables fermionic particles with much larger masses than in the Higgs-free case to form localized states.  
\end{abstract}


\maketitle

\section{Introduction}
The reconciliation of quantum mechanics with general relativity is one of the main outstanding questions of modern physics. Despite the absence of a fully satisfactory theory of quantum gravity, much progress has been made by 
treating the gravitational field classically, most notably perhaps in the prediction of Hawking radiation from black holes \cite{Hawking1975Radiation}. Often, this `semiclassical' approach focuses on the construction of quantum field theories in curved spacetime, but this is limited in its scope. In particular, modeling the full dynamics of general relativity proves challenging, and often the gravitational `back-action' of the matter sector on the spacetime is neglected, or treated as a small perturbation. Consequently, in order to study systems in which the effect of back-action is significant, e.g.\ those with strong self-gravity, an alternative framework is required.

These types of system include gravitationally localized quantum states, in which particles are bound by their gravitational interaction but the state is prevented from collapse by the effects of the uncertainty principle. These are often studied in an approximate framework in which the matter sector is treated not as a quantum field, but instead as a first-quantized wavefunction. In the context of scalar fields, the resulting localized objects are known as ``boson stars" \cite{Kaup1968geon,Feinblum1968boson,Liebling2017BosonStarReview}, which have been proposed as candidates for dark matter \cite{Lee1996bosonStarDarkMatter} as well as black hole mimickers \cite{Torres2000bosonStarBlackHole}. Their fermionic counterparts \cite{Ruffini1969boson,Lee1987solitonStars}, however, have received significantly less attention, due to the added complexity of spin considerations, as well as their limited astrophysical applications. Referred to variously as ``fermion stars", ``Dirac stars", and ``Dirac solitons", these objects could potentially prove useful as models for the microscopic structure of Standard-Model particles.

A major breakthrough in the study of fermionic localized states was made by Finster \textit{et al}.\ \cite{FSY1999original}, who generated the first spherically symmetric, numerical solutions to the coupled Dirac and Einstein equations. The resulting ``particle-like" states, comprising a pair of neutral fermions, are free from singularities, and a branch of solutions has been demonstrated to be stable. Subsequent extensions of their analysis include charged fermions \cite{FSY1999maxwell}, the coupling to an SU(2) Yang--Mills field \cite{FSY2000nonAbelianBound}, and the cases of one \cite{Herdeiro2019bosonDiracProcaSpinning} and many \cite{Bakucz2020powerlaw,Leith2020fermionTrapping,Leith2021excited} fermions.

In this paper we present a hitherto unexplored extension to this framework:\ the addition of a Higgs field. This allows the fermion mass, which in previous analyses has been treated as an input parameter, to be generated dynamically via the Higgs mechanism, as is the case for Standard-Model fermions. By solving the minimally-coupled Einstein, Dirac, and Higgs equations numerically, we show that spherically symmetric particle-like solutions exist for a wide range of parameter values, and are similar in structure to the original Einstein--Dirac states found in ref.~\cite{FSY1999original}. Intriguingly, however, we find a class of solutions in which the total energy of the state (measured by the ADM mass) is no longer proportional to the mass of its constituent fermions. Instead, these two mass scales decouple at strong fermion-Higgs coupling, allowing the ADM mass of the resulting localized state to lie significantly below the fermion mass, and even to decrease as the fermion mass increases.

This paper is organized as follows. In Sec.~\ref{secEOM}, we derive the equations of motion for the Einstein--Dirac--Higgs system, and in Sec.~\ref{secBoundaryConditions} discuss the application of these to gravitationally localized states. In Sec.~\ref{secHiggsDynamics}, we discuss the anticipated dynamics of the Higgs field, before detailing  in Sec.~\ref{secNumericalMethod} the numerical method by which we obtain particle-like solutions. Examples of these solutions are presented in Sec.~\ref{secParticlelikeSolns}, and an analysis of the observed mass-scale separation is given in Sec.~\ref{secMassScales}. We then briefly discuss the issue of stability in Sec.~\ref{secBindingEnergy} before concluding in Sec.~\ref{secDiscussion} with a short summary and discussion.

\section{Equations of motion}\label{secEOM}
We begin by briefly summarizing the derivation of the equations of motion for a pair of neutral fermions interacting via both gravity and a minimally-coupled real scalar Higgs field. Much of this is similar to the Einstein--Dirac case, so we refer the reader to the derivation given in ref.\ \cite{FSY1999original} for further details. Note, however, that precise expressions may differ due to the respective sign conventions employed.

The action for the Einstein--Dirac--Higgs system can be written, using the mostly positive metric signature convention $(-,+,+,+)$, as
\begin{align}
S_{\mathrm{EDH}}=\int \left(\frac{R}{16\pi G}+\mathcal{L}_m \right )\sqrt{-g}\,\mathrm{d}^4x,
\label{eqnAction}
\end{align}
where $g\equiv \mathrm{det}(g_{\mu\nu})$ is the determinant of the spacetime metric $g_{\mu\nu}$, $R$ is the Ricci scalar, $G$ is the gravitational constant, and the Lagrangian density for the gravitational sector has been taken to be of the usual Einstein-Hilbert form. Note that, here and throughout this paper, we employ natural units where $\hbar = c =1$. We also set $G=1$ when generating numerical solutions, and thus all quantities are measured in Planck units. The Lagrangian density for the matter sector can be written as
\begin{align}
\mathcal{L}_m&=\overline\Psi(\slashed{D}-\mu h)\Psi-\frac{1}{2}(\nabla^{\nu}h)(\nabla_\nu h)-V(h),
\end{align}
where $\slashed{D}$ is the Dirac operator in curved spacetime, and $\overline\Psi$ is the usual adjoint spinor. The fermions are minimally coupled (with coupling strength $\mu$) to a Higgs field $h$, which we model as a real scalar field. Hence the fermion mass $\mu h$ becomes a locally varying quantity, set by the local value of $h$. Note that, without loss of generality, we shall henceforth take $\mu>0$.  The Higgs potential $V(h)$ is taken to be of the usual `Mexican hat' form,
\begin{align}
V(h)=\lambda (h^2-v^2)^2\label{eqnMexicanHat},
\end{align}
where the constant $\lambda$ is a positive dimensionless scaling factor, and the stable minima of the potential occur at the vacuum expectation values $h=\pm v$. The mass associated with small displacements around $v$, which we refer to henceforth as the Higgs mass, takes the value:
\begin{equation}
	m_H=2v\sqrt{2\lambda}.
\end{equation}

Extremizing the action (\ref{eqnAction}) with respect to the spinor $\Psi$, the metric $g_{\mu\nu}$, and the Higgs field $h$, gives respectively the Dirac, Einstein, and Higgs equations:
\begin{gather}
(\slashed{D}-\mu h)\Psi=0;\label{eqnDirac}\\
G_{\mu\nu}\equiv R_{\mu\nu}-\frac{1}{2}Rg_{\mu\nu}=8\pi G T_{\mu\nu}\label{eqnEinstein};\\
\nabla_\nu\nabla^\nu h=\mu\overline\Psi\Psi+\frac{\mathrm{d} V}{\mathrm{d} h},\label{eqnHiggs}
\end{gather}
where $T_{\mu\nu}$ is the stress-energy tensor. For simplicity, we seek static, spherically symmetric solutions to this coupled system, corresponding to energy eigenstates. Using the usual spherical polar coordinate system $(t,r,\theta,\phi)$, the metric can be written as
\begin{equation}
g_{\mu\nu}=\mathrm{diag}\left(-\frac{1}{T(r)^2},\frac{1}{A(r)},r^2,r^2\sin^2\theta\right),\label{eqnMetricAnsatz}
\end{equation}
where the forms of the metric fields $T(r)$ and $A(r)$ are to be determined. For the fermionic sector, the simplest case compatible with spherical symmetry is that of two fermions arranged in a singlet state. The appropriate ansatz for the spinor wavefunction in this case is stated in ref.\ \cite{FSY1999original}, and takes the following form:
\begin{equation}
\Psi_a=\frac{\sqrt{T(r)}}{r}\binom{\alpha(r)\chi_a}{-i\beta(r)\sigma^r\chi_a}e^{-i\omega t}.\label{eqnSpinorAnsatz}
\end{equation}
A detailed discussion of the rationale behind this expression can be found in ref.~\cite{Blazquez2020ansatz}. Here, the two fermions, identified by their value of $a\in\{1,2\}$, are assumed to have a common energy $\omega$, with their wavefunctions differing only via the two-component basis vectors $\chi_1=(1,0)^\mathrm{T}$ and $\chi_2=(0,1)^\mathrm{T}$. The radial dependence of the spinor is controlled by the unknown fermion fields $\alpha(r)$ and $\beta(r)$, which can be identified, in the non-relativistic limit, with the fermion and anti-fermion parts of the wavefunction respectively.

Using the ansatzes (\ref{eqnMetricAnsatz}) and (\ref{eqnSpinorAnsatz}), we can reduce the equations of motion (\ref{eqnDirac})--(\ref{eqnHiggs}) to expressions involving the fields $\alpha$, $\beta$, $A$, $T$ and $h$. To do so, we require the form of the Dirac operator in curved spacetime, which in general can be written as $\slashed{D}=i\gamma^\mu\left(\partial_\mu+\Gamma_\mu\right)$, where $\Gamma_\mu$ is the spin connection, and $\gamma^\mu$ are curved-space generalizations of the Dirac gamma matrices, chosen to obey the anticommutation relations $\{\gamma^\mu,\gamma^\nu\}=-2g^{\mu\nu}$. Using the vierbein formalism \cite{Weinberg1972GravitationBook}, one can relate the curved-space gamma matrices to their flat-space counterparts, $\bar\gamma^a$, by the relation $\gamma^{\mu}=e^{\mu}_{\;\;a}\bar\gamma^a$. Considering the metric ansatz (\ref{eqnMetricAnsatz}), we find that the only non-zero vierbein components are $e^{t}_{\;\;t}=T$, $e^{r}_{\;\;r}=\sqrt{A}$ and $e^{\theta}_{\;\;\theta}=e^{\phi}_{\;\;\phi}=1$, and hence the curved-space gamma matrices take the following explicit forms:
\begin{align}
	\gamma^t&=T\bar\gamma^0;\\
	\gamma^r&=\sqrt{A}\left(\bar\gamma^1\sin\theta\cos\phi+\bar\gamma^2\sin\theta\sin\phi+\bar\gamma^3\cos\theta\right);\\
	\gamma^\theta&=\frac{1}{r}\left(\bar\gamma^1\cos\theta\cos\phi+\bar\gamma^2\cos\theta\sin\phi-\bar\gamma^3\sin\theta\right);\\
	\gamma^\phi&=\frac{1}{r\sin\theta}\left(-\bar\gamma^1\sin\phi+\bar\gamma^2\cos\phi\right),
\end{align}
where the flat-space gamma matrices are related to the usual Pauli matrices, as follows:
\begin{align}
	&\bar\gamma^0=\begin{pmatrix}\mathds{1} & 0\\ 0 & -\mathds{1}\end{pmatrix};
	&&\bar\gamma^i=\begin{pmatrix}0 &\sigma^i\\ -\sigma^i & 0\end{pmatrix}.
\end{align}
Utilising this formalism, it can be shown that the Dirac operator in curved spacetime can be written as (see ref.\ \cite{FSY1999original} for details):
\begin{align}
	\slashed{D}&=i\gamma^\mu\partial_\mu+\frac{i}{2}\nabla_\mu\gamma^\mu\label{eqDiracOperator}\\
	&=i \gamma^t \frac{\partial}{\partial t}+i\gamma^r \left( \frac{\partial}{\partial r}+ \frac{1}{r}\left(1-\frac{1}{\sqrt{A}} \right) -\frac{T'}{2T}\right )\notag\\
	&\hspace{50pt}+i\gamma^\theta \frac{\partial}{\partial\theta}+i\gamma^\phi \frac{\partial}{\partial\phi}.
\end{align}
By applying this to the spinor ansatz (\ref{eqnSpinorAnsatz}), the Dirac equation then reduces to the following two coupled differential equations:
\begin{align}
\sqrt{A}\,\alpha'&=+\frac{\alpha}{r}-(\omega T+\mu h )\beta\label{eqnDE1};\\
\sqrt{A}\,\beta'&=-\frac{\beta}{r}+(\omega T-\mu h )\alpha\label{eqnDE2},
\end{align}
where a prime denotes differentiation with respect to $r$. These are identical to those valid in the Einstein--Dirac case, except that the fermion mass is replaced by $\mu h$. Note that here we are considering only states with positive parity. 

To derive an explicit expression for the Einstein equations, we first calculate the stress-energy tensor by varying the matter Lagrangian as per the definition:
\begin{equation}
	T_{\mu\nu}=\frac{-2}{\sqrt{-g}}\frac{\delta}{\delta g^{\mu\nu}}\left(\sqrt{-g}\mathcal{L}_m\right).
\end{equation}
For the fermionic sector, it was shown in ref.\ \cite{FSY1999original} that, for a singlet state, the only contribution to the variation of the Dirac operator is from the first term in (\ref{eqDiracOperator}). To evaluate this, the following identities prove useful:
\begin{align}
	\delta\gamma^{\mu}&=\frac{1}{2}g_{\nu\sigma}\gamma^{\sigma}\delta g^{\mu\nu};\\
	\delta{\sqrt{-g}}&=-\frac{1}{2}\sqrt{-g}\,g_{\mu\nu}\delta g^{\mu\nu};\\
	\delta g_{\sigma\tau}&=-g_{\sigma\mu}g_{\tau\nu}\delta g^{\mu\nu}.
\end{align}
The variation of the Higgs terms in the matter Lagrangian is straightforward, and thus we arrive at the final form of the stress-energy tensor:
\begin{align}
	&T_{\mu\nu}=-\sum_{a=1}^2\Re\left\{\overline\Psi_a\left(i\gamma_\mu\partial_\nu\right)\Psi_a\right\}+(\partial_\mu h)(\partial_\nu h)\notag\\
	&\hspace{70pt}-g_{\mu\nu}\left[\frac{1}{2}(\partial^\sigma h)(\partial_\sigma h)+V(h)\right],
\end{align}
where, as in the Einstein--Dirac case, the contribution from each fermion can simply be added. Using the metric and spinor ansatzes, the non-zero components of the (mixed) stress-energy tensor are therefore found to be:
\begin{align}
	T^{t}_{\;\;t}&=-\frac{2\omega T^2}{r^2}(\alpha^2+\beta^2)-\frac{1}{2}A(h')^2-V(h);\\
	T^{r}_{\;\;r}&=\frac{2T\sqrt{A}}{r^2}\left(\alpha\beta'-\beta\alpha'\right)+\frac{1}{2}A(h')^2-V(h);\\
	T^{\theta}_{\;\;\theta}&=T^{\phi}_{\;\;\phi}=\frac{2T}{r^3}\alpha\beta-\frac{1}{2}A(h')^2-V(h).
\end{align}
Note that only two of these are independent, since the stress-energy tensor is divergenceless, i.e.\ $\nabla_\mu T^{\mu}_{\;\;\nu}=0$.

The components of the Einstein tensor can be obtained from the metric ansatz (\ref{eqnMetricAnsatz}) via the standard sequential procedure of calculating the Christoffel symbols, the Riemann and Ricci tensors, and the Ricci scalar. We find that the only non-zero components are:
\begin{align}
	G^t_{\;\;t}&=\frac{1}{r^2}\left(-1+A+rA'\right);\\
	G^r_{\;\;r}&=\frac{1}{r^2}\left(-1+A-\frac{2rAT'}{T}\right);\\
	G^\theta_{\;\;\theta}&=G^{\phi}_{\;\;\phi}=\frac{A'}{2r}-\frac{A'T'}{2T}+\frac{2A(T')^2}{T^2}-\frac{AT'}{rT}-\frac{AT''}{T}.
\end{align}
Combining these with the stress-energy tensor components above, we find that the Einstein equations reduce to the following two independent differential equations:
\begin{align}
&\frac{1-A}{r^2}-\frac{A'}{r}=8\pi G\bigg(\frac{2\omega}{r^2} T^2(\alpha^2+\beta^2)\notag\\
&\hspace{120pt}\left.+\frac{1}{2}A(h')^2+V(h)\right)\label{eqnEE1};\\
&\frac{1-A}{r^2}+\frac{2AT'}{rT}=8\pi G\bigg(\frac{2T\sqrt{A}}{r^2}(\beta\alpha'-\alpha\beta')\notag\\
&\left.\hspace{120pt}-\frac{1}{2}A(h')^2+V(h)\right).\label{eqnEE2}
\end{align}
The bracketed terms in these two equations are respectively the energy density and the radial pressure in the matter sector. 

Finally, the Higgs equation (\ref{eqnHiggs}) can be rewritten, using the metric and spinor ansatzes, as:
\begin{equation}
Ah''-A\left(\frac{T'}{T}-\frac{A'}{2A}-\frac{2}{r}\right)h'=\frac{2\mu}{r^2}T(\alpha^2-\beta^2)+\frac{\mathrm{d}V}{\mathrm{d}h}\label{eqnH}.
\end{equation}
Together, equations (\ref{eqnDE1}), (\ref{eqnDE2}), (\ref{eqnEE1}), (\ref{eqnEE2}) and (\ref{eqnH}) constitute a coupled set of five differential equations, for the five unknown fields $\alpha$, $\beta$, $A$, $T$ and $h$. Within the semi-classical framework considered here, these equations fully define the behavior of a pair of static, neutral fermions interacting via gravity and a minimally-coupled Higgs field.

\section{Boundary conditions}\label{secBoundaryConditions}
Our aim is to seek particle-like solutions to the above equations of motion, representing gravitationally localized states. We therefore require the following boundary conditions. First, the metric should be asymptotically flat, implying $A(r),T(r)\rightarrow 1$ as $r\rightarrow\infty$. Second, the fermion wavefunction should be correctly normalized, i.e.~the inner product $(\Psi|\Psi)=1$. Using (\ref{eqnSpinorAnsatz}), this can be rewritten as
\begin{equation}
4\pi\int_0^\infty\frac{T}{\sqrt{A}} \left(\alpha^2+\beta^2\right)\mathrm{d}r=1.\label{eqnNorm}
\end{equation}
In order for asymptotic flatness to be satisfied, the energy-density contribution from the Higgs field must vanish at large $r$. Considering (\ref{eqnEE1}) and (\ref{eqnEE2}), this is achieved when $h'=0$ and $V(h)=0$, implying $h\rightarrow \pm v$ as $r\rightarrow\infty$, i.e.\ the Higgs field should asymptote to one of its two possible vacuum expectation values.

In addition, we are required to specify boundary conditions at $r=0$. As in the Einstein--Dirac case, it is possible to obtain a small-$r$ expansion which guarantees that the particle-like states are free from central singularities. This can be written as follows:
\begin{align}
	\alpha(r)&=\alpha_1 r+...\label{eqSmallrStart}\\
	\beta(r)&=\frac{1}{3}\alpha_1(\omega T_0-\mu h_0)r^2+...\\
	T(r)&=T_0+\frac{4\pi G}{3}T_0\left(V(h_0)+\alpha_1^2T_0(\mu h_0-\omega T_0)\right)r^2+... \\
	A(r)&=1-\frac{8\pi G}{3}\left(V(h_0)+2\omega \alpha_1^2 T_0^2 \right)r^2+...\\
	h(r)&=h_0+\frac{1}{3}\left(2\lambda h_0(h_0^2-v^2)+\mu\alpha_1^2 T_0\right )r^2+...\label{eqSmallrEnd},
\end{align}
where the coefficients $\alpha_1$, $T_0$ and $h_0$ are unconstrained. We note that there is no guarantee that this is the unique expansion for which non-singular states can be generated, but it is certainly that which will produce solutions most similar to those found in the Einstein--Dirac case.

A cursory glance at the above expansion, along with the equations of motion, suggests that $\mu$, $\lambda$, $v$, $\omega$, $\alpha_1$, $T_0$ and $h_0$ are all free parameters within the theory. This is not the case however:\ imposing normalization and asymptotic flatness together removes a total of four degrees of freedom, and thus only three of these may be freely specified. As in the Einstein--Dirac case, one of these parameters is the central redshift $z\equiv T(0)-1$, which can be employed as a measure of how relativistic a state is, with $z\approx 1$ marking the crossover from non-relativistic to relativistic. For the two remaining free parameters, we could choose $m_f=\mu v$ and $m_H=2v\sqrt{2\lambda}$, but it turns out to be significantly more computationally efficient to choose instead $v$ and $\xi$, where $\xi$ is the Higgs to (asymptotic) fermion mass ratio:
\begin{equation}
	\xi\equiv \frac{m_H}{m_f}=\frac{2\sqrt{2\lambda}}{\mu}.
\end{equation}
We find that each choice of $\{v,\xi\}$ then defines a one-parameter family of solutions where, as for the Einstein--Dirac case, the value of the central redshift $z$ uniquely identifies states within each family. The four remaining parameters ($\mu$, $\omega$, $\alpha_1$ and $h_0$) will take values dependent on the state in question. We note that this is by no means the only parametrization possible, and an example of an alternative can be found in Appendix \ref{appPar}.

\section{Higgs field dynamics}\label{secHiggsDynamics}

Before presenting our numerical results, we first outline the expected behavior of the Higgs field for particle-like states, employing a similar rationale to Schl{\"o}gel \textit{et al.}\ \cite{Schlogel2014Higgs}. Consider first the situation outside the localized fermion source, where the fermion fields $\alpha$ and $\beta$ are negligible. Let us also temporarily introduce a time-dependence to the Higgs field. In this case, the Higgs equation can be written as
\begin{equation}
T^2\ddot h-Ah''+A\left(\frac{T'}{T}-\frac{A'}{2A}-\frac{2}{r} \right )h'=-\frac{\mathrm{d}V}{\mathrm{d}h},
\end{equation}
where the dot denotes a time-derivative. Note the sign difference between the $\ddot h$ and $h''$ terms:\ this implies that the two dynamically stable minima in the Higgs potential ($h=\pm v$) are unstable maxima from the point of view of spatial variations \cite{colemanbook}. Hence, in the static case that we are considering here, the Higgs potential is effectively inverted compared to its usual form.

\begin{figure}
	\includegraphics{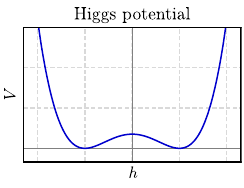}\hspace{3pt}
	\includegraphics{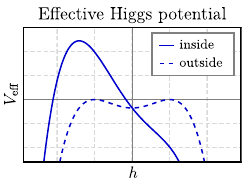}
	\caption{Left: The intrinsic Mexican-hat Higgs potential $V(h)$, with two stable minima at $h=\pm v$. Right: The (inverted) effective Higgs potential $V_{\mathrm{eff}}(h)$, which exhibits a tilt inside the localized fermion source.}
	\label{figPotential}
\end{figure}

At positions within the fermion source, the coupling to the fermions introduces an additional term on the right-hand side of the Higgs equation. Combining this with the Higgs potential $V(h)$, we can rewrite the Higgs equation as $\nabla_\mu\nabla^\mu h=-\partial V_{\mathrm{eff}}/\partial h$, where we have defined an effective potential that takes the form
\begin{equation}
V_{\mathrm{eff}}(h)=-\lambda (h^2-v^2)^2-\frac{2\mu}{r^2}T(\alpha^2-\beta^2)h.
\label{eqnEffectivePotential}
\end{equation}
Hence the fermionic term introduces a `tilt' to the intrinsic Mexican-hat Higgs potential via a term that is linear in $h$; this is illustrated schematically in Fig.~\ref{figPotential}.

We shall consider here only states in which $\alpha$ and $\beta$ are nodeless (ground states), and only those with positive parity. As in the Einstein--Dirac case, we find that these correspond to states with positive (asymptotic) fermion mass. Consequently, $m_f=\mu v$ must be positive, implying that the Higgs field outside the fermion source should asymptote to $h=+v$. In addition, since $\alpha$ is the dominant fermion field, the cumulative tilt to the Higgs potential must always be in the direction indicated in Fig.~\ref{figPotential}, implying $h<v$ within the fermion source.

\section{Numerical method}\label{secNumericalMethod}

We now outline the numerical method via which particle-like states of the Einstein--Dirac--Higgs equations can be obtained. First, in order to deal with the conditions of asymptotic flatness and normalization, we implement a `rescaling' procedure similar to that employed in ref.\ \cite{FSY1999original}. This relies on the fact that, for localized states (where $\alpha,\beta\rightarrow 0$ and $h\rightarrow v$ at large $r$), the equations of motion automatically imply $A\rightarrow 1$ and $T\rightarrow\mathrm{const.}$ as $r\rightarrow\infty$, and the normalization integral will evaluate to a constant. Thus, it is sufficient to generate first an `unscaled' solution (by temporarily specifying values for $T_0$ and $\mu$), for which the fermion fields decay at large $r$ and the Higgs field asymptotes to $v$, but which is not correctly normalized nor asymptotically flat. The true, physical solution can then be obtained by simply rescaling the fields such that both $T(\infty)$ and the normalization integral are equal to 1. 

More formally, having generated an unscaled solution (denoted by a tilde), for which $\tilde T_0=1$ and $\tilde\mu=1/v$, we define
\begin{gather}
	\tau=\lim_{r\rightarrow\infty}\tilde T(r);\\
	\chi^2=4\pi\int_0^\infty \left(\tilde\alpha^2+\tilde\beta^2\right)\tilde T\tilde A^{-1/2}\,\mathrm{d}r,
\end{gather}
and then rescale the fields and parameters as follows, to obtain the physical solution:
\begin{align}
	\alpha(r)&=\sqrt{\frac{\tau}{\chi}}\,\tilde \alpha(\chi  r);&
	\beta(r)&= \sqrt{\frac{\tau}{\chi}}\,\tilde\beta(\chi r);\notag\\
	T(r)&= \frac{1}{\tau} \tilde T(\chi r);&
	A(r)&= \tilde A(\chi r);\notag\\
	h(r)&=  \tilde h(\chi r);&
	\omega&=  \chi\tau\tilde\omega;\notag\\
	\mu&= \chi\tilde\mu;&
	\lambda&= \chi^2\tilde\lambda.
\end{align}
Note that the value of the Higgs field itself, and thus its vacuum expectation value, remains unaltered under the rescaling procedure.

We have now reduced the problem to one of obtaining unscaled solutions to the equations of motion. These can be generated by tuning the values of $\tilde\omega$ and $h_0$ such that the fermion fields become normalizable (decay sufficiently rapidly at large $r$) and the Higgs field asymptotes to $v$. Due to the coupling between the fields, however, $\tilde\omega$ and $h_0$ cannot be sought independently, and thus a two-parameter shooting procedure is required. Fortunately, it is possible to implement this sequentially by first choosing a value for $h_0$, performing a simple binary chop to find the ground-state value of $\tilde\omega$ for which the fermions become normalizable, and then noting the behavior of the Higgs field at large $r$. If $h(\infty)<v$, then $h_0$ should be increased; if $h(\infty)>v$, then $h_0$ should be decreased. By iterating this procedure, it is possible to force $h(\infty)$ to its vacuum expectation value, while ensuring that the fermions remain normalizable. Note that this works well only for solutions that are fermion-dominated ($\xi<2$); for Higgs-dominated states, the shooting order should be reversed such that a binary chop in $h_0$ is performed at chosen values of $\tilde\omega$.

There is one further complication that arises when numerically generating solutions. As mentioned, each physical ground-state solution can be uniquely identified by the values of three parameters: $\xi$, $v$ and $z$. In the unscaled system, however, the role of the central redshift is taken by the parameter $\tilde\alpha_1$, which, unlike in the Einstein--Dirac system, is not necessarily guaranteed to be in one-to-one correspondence with $z$. Indeed, for small values of $\xi$ and $v$, we find that there are regions in which multiple (ground-state) solutions can be found with the same value of $\tilde\alpha_1$, but upon rescaling their values of $z$ are found to differ. This property further complicates the solution-finding procedure, since we must ensure that all solutions for a particular value of $\tilde\alpha_1$ have been identified. To achieve this, we perform an initial coarse-grained sweep of the solution space by varying $h_0$ over the region $[-2v,v]$, from which we can ascertain the presence of additional solutions by noting changes in the asymptotic behavior of the Higgs field.

\section{Particle-like solutions}\label{secParticlelikeSolns}
We now present numerical results illustrating the structure of the particle-like states present in the Einstein--Dirac--Higgs system. These are generated via the method outlined in the preceding section, using Mathematica's built-in differential equation solver, NDSolve, with a minimum accuracy of 8 digits. We shall here discuss only ground-state solutions (i.e.\ those in which $\alpha$ and $\beta$ are nodeless), although an example of an excited state can be found in Appendix \ref{appExcited}.

\begin{figure*}
	\includegraphics{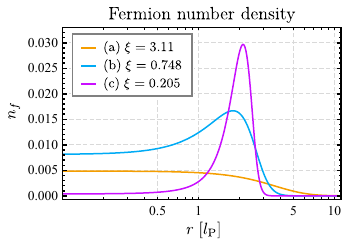}\hspace{7pt}
	\includegraphics{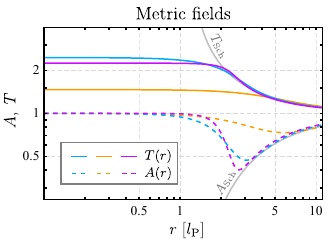}\hspace{7pt}	
	\includegraphics{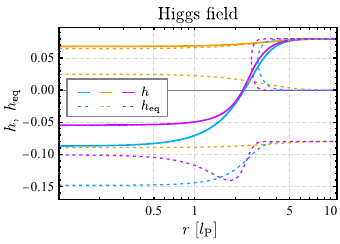}
	\caption{Fermion number density, metric and Higgs field profiles for three particle-like solutions of the Einstein--Dirac--Higgs system, corresponding to the three states labeled in Fig.~\ref{figMassScales}. These have equal values of $v=0.08$, but different values of the Higgs-to-fermion mass ratio $\xi$ and of the central redshift $z$. Included alongside the metric fields are the Schwarzschild profiles corresponding to a point source with ADM mass equal to that of solution (c). The dashed lines included alongside the Higgs field profiles track the values of $h$ for which $\mathrm{d}V_{\mathrm{eff}}/\mathrm{d}h=0$, i.e.\ the equilibrium positions of the effective Higgs potential. Within the fermion source, these depart from their asymptotic values, $\{0,\pm v\}$, due to the fermion tilt. The values of the fermion mass, ADM mass and fermion-Higgs coupling strength for the three states are as follows --- (a): $\{m_f=0.558,M=1.08,\mu=6.97\}$, (b): $\{m_f=1.00,M=0.946,\mu=12.6\}$, and (c): $\{m_f=3.68,M=0.899,\mu=46.0\}$. Full details of these and other parameters associated with the solutions are given in Appendix \ref{appData}.}
	\label{figIndividualSolns}
\end{figure*}

We have been successful in generating solutions with parameter values ranging approximately from $\{v,\xi\}=\{0.07,0.03\}$--$\{10,30\}$. The upper limits on this arise from issues concerning numerical precision, but the reason behind the lower limits is less clear. In particular, we have been unable to obtain solutions below $v=0.07$ for any value of $\xi$. We suspect that this is associated with the failure of our numerical method, since we observe indications of a further degree of multivaluedness (similar to that detailed in ref.\ \cite{Leith2021excited}) appearing at very small values of $v$. In addition, no parameters appear to become singular as $v=0.07$ is approached, and therefore we tentatively conclude that solutions below this threshold do indeed exist, but we are unable to generate them.

Examples of three particle-like solutions are shown in Fig.~\ref{figIndividualSolns}. These have a common value of $v=0.08$, but differ in their values of $\xi$ and $z$. Solution (a) is Higgs-dominated ($\xi>2$), while solutions (b) and (c) are fermion-dominated ($\xi<2$). Plotted on the left are the radial profiles of the fermion number density $n_f(r)$, defined as
\begin{equation}
n_f=\frac{2T}{r^2}(\alpha^2+\beta^2),
\end{equation}
which takes considerably different forms for the three states. For all three, $n_f$ decays exponentially at large $r$ (consistent with the notion of a localized state), but in solution (a), the peak occurs at $r=0$ (as in the Einstein--Dirac case), whereas for solutions (b) and (c) it is shifted significantly outwards in radius.

The metric fields are singularity free, with $T(r)$ decreasing\ monotonically from a central maximum, while $A(r)\leq 1$ throughout. Outside the localized fermion source, the metric fields approach the standard Schwarzschild form, for which $T_{\mathrm{Sch}}^{-2}=A_{\mathrm{Sch}}=1-2GM/r$. This allows us to identify an Arnowitt-Deser-Misner (ADM) mass $M$, which provides a measure of the total energy of the localized state. 

With regard to the Higgs field, in all three solutions this rises from a constant central value before asymptoting towards $v=0.08$. In solution (a), the effect of the fermion tilt is such that the Higgs field deviates only slightly from its vacuum expectation value, and hence its properties are similar to that of an Einstein--Dirac state (where $h$ is pinned at $v$). For solutions (b) and (c), however, the fermion tilt is large enough that the Higgs field becomes negative at small $r$, resulting in the local fermion mass becoming negative within the central regions of the fermion source.

\begin{figure}[b]
	\includegraphics{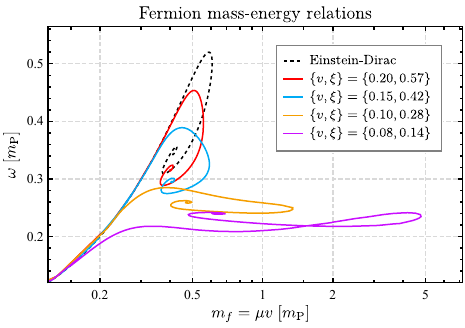}
	\caption{Fermion mass-energy relations for the families of Einstein--Dirac--Higgs states defined by the Higgs vacuum expectation values $v$ and Higgs-to-fermion mass ratios $\xi$ indicated. Included also is the corresponding curve for the Einstein--Dirac case. Note the increase in the maximum fermion mass as $v$ and $\xi$ are decreased.}
	\label{figSpirals}
\end{figure}

The extent to which the Einstein--Dirac--Higgs states differ from their Einstein--Dirac counterparts can be more clearly illustrated by considering the families of solutions defined by the values of the Higgs vacuum expectation value $v$ and the Higgs-to-fermion mass ratio $\xi$. This is shown in Fig.~\ref{figSpirals}, where we plot the fermion energy $\omega$ as a function of the asymptotic fermion mass $\mu v$, for a selection of families, alongside the known Einstein--Dirac relation. As can be seen, the spiraling behavior is preserved in the Higgs case, with each curve initially approximating the non-relativistic relation $\omega=m_f$ before spiraling towards an infinite-redshift solution (see Appendix \ref{appPL}). In the relativistic regime, however, the families with smaller values of $v$ and $\xi$ begin to deviate significantly from the Einstein--Dirac relation, and, in particular, the maximum allowable fermion mass increases substantially. 

\section{Mass-scale separation}\label{secMassScales}
In this section, we demonstrate the existence of a somewhat unexpected phenomenon in the Einstein--Dirac--Higgs system: a mass-scale separation occurs at strong fermion-Higgs coupling, with the ADM mass of a state no longer being proportional to the mass of its constituent fermions. As an example, consider the mass scales of the three states shown in Fig.~\ref{figIndividualSolns}. Despite their similar ADM masses ($1.08$, $0.946$ and $0.899$ respectively), they have significantly different asymptotic fermion masses ($0.558$, $1.00$ and $3.68$ respectively), as a consequence of their differing $\mu$ values. Why, one might wonder, has the ADM mass not increased in proportion with the fermion mass for the states in which the fermion-Higgs coupling is strong, i.e.\ when the value of $\mu$ is large? 

Before exploring potential reasons for this, we first illustrate the effect more clearly by analyzing the families of states defined by $v$ and $\xi$, and in particular the properties of the maximally-bound state in each family. Note that, as in the Einstein--Dirac case, we find that the most bound state in each family appear always to correspond to that of maximum fermion mass. We have shown previously in Fig.~\ref{figSpirals} that the value of the maximum fermion mass shows a general increase with both decreasing $v$ and $\xi$, but what is the corresponding change in the ADM mass? 

\begin{figure}
	\includegraphics{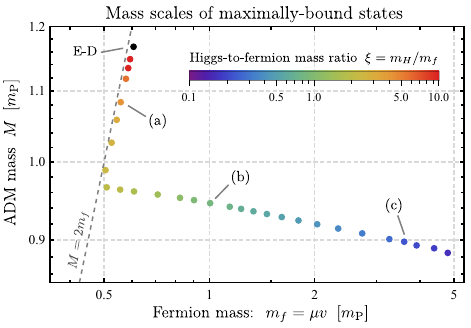}
	\caption{The ADM mass of a localized, two-fermion state as a function of the mass of the constituent fermions. Plotted are the most bound states (those with maximum possible fermion mass) for a fixed value of the Higgs vacuum expectation value, $v=0.08$, and for various values of the Higgs-to-fermion mass ratio $\xi$. The three labeled states correspond to those shown in Fig.~\ref{figIndividualSolns}. For solutions with large values of $\xi$, the ADM mass of the state is in direct proportion to the fermion mass, and approaches the Einstein--Dirac relation found by Finster \textit{et al}.~\cite{FSY1999original} as $m_H \rightarrow \infty$. There is, however, a separate class of states, with lower values of $\xi$, for which the ADM mass becomes parametrically smaller than the sum of the masses of the constituent fermions.  This allows states with ever higher fermion mass to be formed as $m_H$ is decreased.}
	\label{figMassScales}
\end{figure}

To answer this, consider Fig.~\ref{figMassScales}, in which we have isolated the maximally-bound states for a selection of families with $v=0.08$ but differing values of $\xi$, and plotted the ADM mass of each as a function of fermion mass. This clearly indicates a decoupling of mass scales at $\xi\approx 2$, with states above this threshold exhibiting an ADM mass that is approximately twice the fermion mass (as in the Einstein--Dirac case), while states with $\xi<2$ depart significantly from this expectation. Indeed, in the most extreme case shown here, the total fermion mass is over ten times larger than the ADM mass of the state.

This increase in fermion mass appears to be driven primarily by a corresponding increase in the fermion-Higgs coupling strength $\mu$. When we generate solutions, however, $\mu$ is not an input parameter; instead we specify the values of $v$ and the Higgs-to-fermion mass ratio $\xi$. What ranges of $v$ and $\xi$, then, correspond to strong coupling? This can be answered by exploring the phase space spanned by $v$ and $\xi$. To do so, we require a reference solution that allows us to compare properties across the various families of states. Ideally, we would choose the maximally-bound state, in line with our earlier analysis, but unfortunately this proves computationally unfeasible, since it requires a large portion of each spiraling family to be generated for every value of $\xi$ and $v$. Instead, we utilise the fact that the \textit{minimally}-bound state in each family occurs at an approximately constant value of $\tilde\alpha_1=0.25$, and can therefore be more readily employed as a reference solution.

\begin{figure*}
	\includegraphics{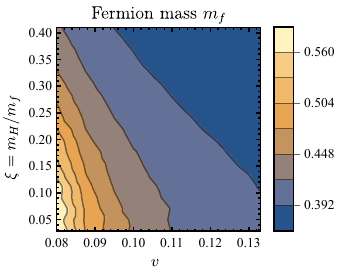}\hspace{7pt}
	\includegraphics{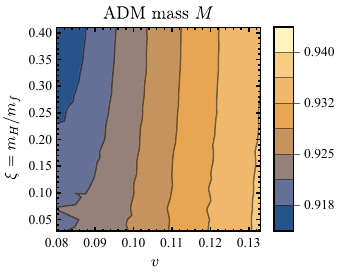}\hspace{7pt}
	\includegraphics{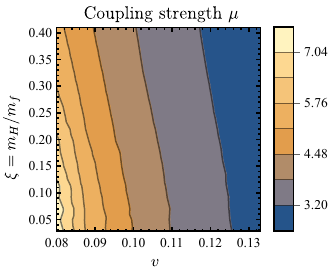}	
	\caption{Contour plots illustrating the behavior of the fermion mass $m_f$, ADM mass $M$ and fermion-Higgs coupling strength $\mu$, as a function of the parameters $\xi$ and $v$, for the least-bound states in each family. Note that the contours become less smooth at smaller values of $\xi$ where errors arising from our numerics are largest.}
	\label{figPhase}
\end{figure*}

The results of this analysis are illustrated in Fig.~\ref{figPhase}, where we plot the fermion mass, ADM mass and fermion-Higgs coupling strength as a function of both $v$ and $\xi$, with the data corresponding to the minimally-bound state in each family. From these, we observe that smaller values of both $v$ and $\xi$ correspond to larger values of $\mu$, and in addition that regions of strong coupling roughly equate to regions of large fermion mass. This is consistent with the behavior shown previously in Fig.~\ref{figSpirals}. Furthermore, the value of the ADM mass exhibits little variation over the entire range of $v$ and $\xi$ plotted, whereas the fermion mass increases significantly by comparison as $v$ and $\xi$ decrease. This confirms that indeed the mass-scale separation is most prominent at small values of $v$ and $\xi$ (and hence at strong coupling). We note that it is not surprising that small $\xi$ implies large $m_f$ (and $\mu$), since $\xi\sim 1/\mu$, but it is unclear why small values of $v$ imply the same.

Although it is clear from the above analysis that the observed mass-scale separation is driven primarily by an increase in the fermion-Higgs coupling strength, the precise mechanism through which this is achieved remains unclear. In particular, we can provide no satisfactory answer as to why the ADM mass of states at strong coupling does not increase in proportion to the constituent fermion mass. Some progress can be made, however, by expressing the ADM mass using the Komar integral \cite{WaldGR}, which reduces to:
\begin{align}
&M=4\omega-8\pi\int_0^\infty\frac{\mathrm{d}r}{\sqrt{A}}\left[
\mu h(\alpha^2-\beta^2)+\frac{r^2}{T}V(h)\right].
\end{align}
For states with large values of $\mu$, the Higgs field is initially negative due to the strong fermion tilt (see Fig.~\ref{figIndividualSolns}), and hence the first term in the integral switches sign between the inner and outer regions of the fermion source. Its overall contribution is therefore negligible, and thus the large value of $\mu$ does not directly affect the ADM mass.

This is not a complete explanation since we would expect the fermion energy $\omega$ also to scale with the fermion mass. Instead, we find the reverse is true; $\omega$ tends to be lower for states in which the fermion mass is large. Without knowing what precisely affects the value of $\omega$, it is difficult to put forward an explanation for this. One suggestion may be that the change in the fermion density profiles (itself a consequence of the strong fermion-Higgs coupling) prevents $\omega$ from increasing. Also potentially related is the observed disparity between the radial decay scales of the fermion and Higgs fields (see ref.\ \cite{Leith2022thesis} for details), which occurs only for $\xi<2$. The link between this and the value of $\omega$, however, is not clear.

\section{Binding energy \& stability}\label{secBindingEnergy}

\begin{figure}
	\includegraphics{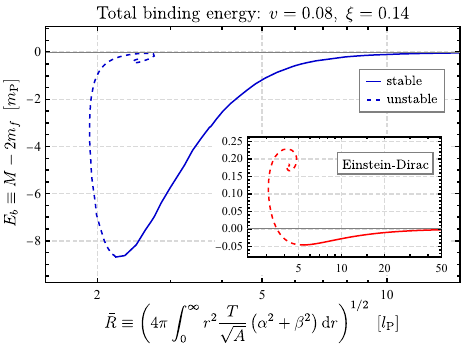}
	\caption{Binding energy as a function of (rms) radius $\bar R$ for the family of localized states with $v=0.08$ and $\xi=0.14$. In the Einstein--Dirac case (inset), only a portion of the curve contains solutions that are bound (those with negative binding energy), but with the addition of the Higgs field it is possible for all solutions in a family, even the highly-relativistic states located near the center of the spiral, to become bound.}
	\label{figBindingEnergyAllBelow}
\end{figure}

The decoupling of mass scales also affects the binding energy of states, defined as $E_b=M-2m_f$, i.e.\ the difference between the energy of the state and that of two individual delocalized fermions. With this definition, a state is considered bound if it has a negative value of $E_b$ (energy is required to break it apart). As we have shown, it is possible for the fermion mass to far outweigh the ADM mass at strong coupling, and as such these states are highly bound. Indeed, in contrast to the Einstein--Dirac case, it is even possible for entire families of solutions to become bound. An example of this is shown in Fig.~\ref{figBindingEnergyAllBelow}.

It is important to note, however, that negative binding energy does not necessarily imply stability. Indeed, the only conclusion that can be drawn from a binding energy analysis is whether the state will remain spatially localized upon (infinitesimal) perturbation. Although this of course includes the case of a stable state, it also encompasses the possibility of gravitational collapse, e.g.~to a black-hole type object, or indeed decay to another localized state, both of which have been observed in dynamical simulations of Dirac stars \cite{Daka2019diracStarEvolution}.

A full stability analysis of the Einstein--Dirac--Higgs states presented here is beyond the scope of this paper, but we can gain some insight by invoking comparisons with the Einstein--Dirac case, the stability of which has been examined, both analytically and numerically, in refs.~\cite{FSY1999original} and \cite{Daka2019diracStarEvolution}. In particular, Finster \textit{et al.}~\cite{FSY1999original} demonstrate, via Conley index theory, that Einstein--Dirac states with redshifts lower than that of the state of maximum fermion mass are stable, while those above this threshold are unstable. Their argument relies on the spiraling nature of the family of solutions, in particular that there is a continuous connection to the non-relativistic regime, where solutions are known to be stable, as well as the existence of a well-defined quantity (in this case the fermion mass) that can be used as a bifurcation parameter.

The same argument can be applied in the Einstein--Dirac--Higgs case. As evidenced in Fig.~\ref{figSpirals}, the families of states exhibit a similar spiraling behavior. Moreover, due to the choice of our parametrization, the value of the Higgs vacuum expectation value $v$ is constant within each family, and hence it is possible for the (asymptotic) fermion mass $m_f=\mu v$ to serve as a bifurcation parameter. In addition, each state is continuously connected to the non-relativistic regime, where solutions involve a Higgs field that deviates from its vacuum expectation value by an ever smaller amount as $m_f\to 0$. Since these low-mass, non-relativistic states so closely resemble their Einstein--Dirac counterparts, it seems reasonable to expect that they should be stable, and thus a continuous connection to a stable state can be established. Finally, one might be concerned that the existence of multiple families of solutions results in additional decay pathways compared to the Einstein--Dirac case. However, since the properties of the Higgs field and Higgs potential (governed by $m_H$, $\lambda$ and $v$) are global features, it seems reasonable to consider only dynamical processes in which these quantities remain fixed, and thus transitions between families are excluded.

We therefore conclude that the same stability criterion that applies in the Einstein--Dirac case should equally apply here, i.e.~each family of states should contain a stability transition point corresponding to the respective state of maximum fermion mass. We note that this behavior is similar to that exhibited by other gravitationally localized objects, such as boson stars \cite{Gleiser1989Stability}. The expected stability regimes for the $\{v,\xi\}=\{0.08,0.14\}$ family are indicated in Fig.~\ref{figBindingEnergyAllBelow}. We note that, for all families of states, the stability transition point appears to coincide with the minimum of the binding energy $E_b$, but not necessarily the \textit{global} maximum of the ADM mass. Instead, the minimum in $E_b$ only coincides with a \textit{local} maximum of $M$ for families where $\xi$ and $v$ are small. This differs from the Einstein--Dirac case, where the global extrema of $m$, $M$ and $E_b$ all coincide, and is another consequence of the mass-scale separation that occurs when the Higgs field is introduced.

The above argument establishes the existence of a stable branch of Einstein--Dirac--Higgs states for each value of $\xi$ and $v$. We therefore conclude that the inclusion of the Higgs field allows stable bound states to exist in which the constituent fermions are significantly more massive than allowed in the Einstein--Dirac system. These correspond to states in which the mass-scale separation is prominent.

\section{Summary \& Discussion}\label{secDiscussion}
In this paper, we have constructed gravitationally localized solutions to the minimally-coupled Einstein--Dirac--Higgs system, and have shown that the resulting particle-like states are well behaved and free from singularities. Somewhat unexpectedly, at strong fermion-Higgs coupling, we find that the ADM mass appears to become parametrically smaller than the mass of the constituent fermions, allowing fermions of much larger mass than in the Einstein--Dirac case to form localized states. 

The implications of this mass-scale separation are somewhat intriguing. In particular, for states in which it is a dominant feature, the disparity between mass scales implies that much of the mass of the constituent fermions is `hidden' from an external observer (at least from a gravitational point of view). For the solutions presented in this paper, the largest disparity observed is of approximately a factor of ten, but it may be possible for even more extreme situations to occur. For example, we have been unable to determine whether the downward trend shown in Fig.~\ref{figMassScales} continues at even lower values of $\xi$ (due to numerical difficulties), but if an extrapolation can be trusted, it would imply that states may exist which contain ultra-high-mass fermions, but whose external gravitational mass is comparatively negligible.

Finally, we note the existence of a series of papers by Dzhunushaliev \textit{et al}. \cite{Dzhunushaliev2019DiracStarsNonLinear,Dzhunushaliev2019DiracStarMaxwellProca,Dzhunushaliev2020DiracStarYangMills}, which appear to report an effect similar to that discussed here, although in the context of including additional non-linearities and/or Proca fields within the Einstein--Dirac system. Our analysis differs from these, however, in a number of important ways. First, the models considered by Dzhunushaliev \textit{et al}.\ are restricted to a contact interaction between the fermions; in our model, by contrast, fermion-fermion interactions are mediated by the Higgs field and thus occur at all spatiotemporal separations consistent with causality. Second, their analysis is performed at a purely classical level, i.e.\ without imposing the normalization of the spinor wavefunction, and as a result the fermion mass exists as a free parameter within the system. Nonetheless, it is certainly true that a branch of solutions does exist in which the ADM mass of states fails to scale with the mass of the constituent fermions, but the mechanism through which this is achieved differs significantly from that discussed here.

To conclude, we emphasize that the analysis contained within this paper is restricted to a semi-classical approximation, and its applicability in a more rigorous quantum context is unclear. Nevertheless, our results provide an illustrative study of how fermionic objects of a finite extent may be expected to interact with a Higgs field, within the framework of general relativity.

\begin{acknowledgments}
PEDL acknowledges funding from a St Leonards
scholarship from the University of St Andrews and from
UKRI under EPSRC Grant No.~EP/R513337/1. ADL is grateful for financial support from the EPSRC (UK) under grant number EP/L505079/1.
\end{acknowledgments}

\appendix

\section{Data for figures}\label{appData}

\setlength{\tabcolsep}{4.5pt}
\begin{table*}[t]
	\caption{Parameter values for the individual localized states shown in figures throughout the paper.}
	\begin{tabular}{ |c|ccccccccc| } 
		\hline
		Figure & $v$ & $z$ & $\mu$ & $\lambda$ & $\omega$ & $\alpha_1$ & $h_0$ & $M$ & $\bar R$ \\ 
		\hline
		\ref{figIndividualSolns}(a) & 0.08 & 0.46440042 & 6.9726065 & 58.826861 & 0.46654990 & 0.040731809 & 0.068129517 & 1.0838776 & 5.6178928 \\ 
		\ref{figIndividualSolns}(b) & 0.08 & 1.4559839 & 12.558979 & 11.036032 & 0.25207591 & 0.040663327 & $-0.087130486$ & 0.94600990 & 2.4967371 \\
		\ref{figIndividualSolns}(c) &   0.08 & 1.2450824 & 46.018093 & 11.130976 & 0.23545904 & 0.0090303129 & $-0.054588173$ & 0.89889341 & 2.2349828 \\
		\ref{figPLZ} & 0.1 & 34176.148 & 4.7236279 & 0.22312660 & 0.25907640 & 0.078863525 & $-0.082817471$ & 0.94212579 & 2.9007476 \\
		\ref{figPLSoln} & 0.1 & $\infty$  & 4.6770234 & 0.21874548 & 0.25955448 & -- & $-0.081575838$ & 0.94171974 & 2.9083848 \\
		\ref{figExcited} & 0.1 & 1.1720432 & 11.607477 & 1.3473353 & 0.98587504 & 0.077537351 & 0.024036623 & 2.3021427 & 13.839702 \\
		\hline
	\end{tabular}
\end{table*}

Throughout this paper, we have presented examples of individual localized states in the Einstein--Dirac--Higgs system. Various parameter values associated with these solutions are presented in Table I, from which the states could be reconstructed, if desired. Note that the values of the Higgs mass and asymptotic fermion mass can be obtained from these using the relations $m_f=\mu v$ and $m_H=2v\sqrt{2\lambda}$.

\section{Power-law solution \& infinite-redshift states}\label{appPL}

For the Einstein--Dirac system, it was shown in ref.\ \cite{Bakucz2020powerlaw} that the radial structure of localized states can be understood in terms of distinct zones. In particular, for high-redshift solutions, there exists a `power-law' zone where the solution approximates that of the massless Einstein--Dirac equations, for which all fields have a simple power-law dependence on $r$. It was also demonstrated that the infinite-redshift solution located at the center of the spiraling curves contains a power-law zone that extends all the way to $r=0$. Here, we show that similar properties exist for the states in the Einstein--Dirac--Higgs system. 

First, we derive the analog of the massless power-law solution. In the context of the Einstein--Dirac--Higgs system, this requires $\mu h\ll \omega T$, i.e.\ that the local fermion mass is negligible compared to the local fermion energy. Since this occurs in regions in which the solution is highly relativistic, we expect the energy density from the fermions to dominate over the contribution from the Higgs field. Then the Dirac and Einstein equations reduce to:
\begin{align}
	\sqrt{A}\,\alpha'&=+\frac{\alpha}{r}-\omega T\beta;\\
	\sqrt{A}\,\beta'&=-\frac{\beta}{r}+\omega T\alpha;\\
	1-A-rA'&=16\pi G\omega T^2(\alpha^2+\beta^2);\\
	1-A+\frac{2rAT'}{T}&=16\pi GT\sqrt{A}(\beta\alpha'-\alpha\beta').
\end{align}
This is precisely the massless Einstein--Dirac system; thus the power-law zone is a region in which the Higgs field has no effect on either the metric or the distribution of the fermion source. From ref.\ \cite{Bakucz2020powerlaw}, the solution to the above system is
\begin{align}
	&\alpha(r)=\alpha_p r;
	&&\beta(r)=\beta_p r;\notag\\
	&A(r)=\frac{1}{3};
	&&T(r)=\sqrt{\frac{2}{3}}\,\omega r^{-1}\label{eqPL1},
\end{align}
where
\begin{align}
	\alpha_p&=3^{1/4}\sqrt{\frac{\omega}{48\pi G\left(\sqrt{3}-1\right)}};\\
	\beta_p&=3^{1/4}\sqrt{\frac{\omega}{48\pi G\left(\sqrt{3}+1\right)}}.
\end{align}
The behavior of the Higgs field in the power-law zone is governed by the Higgs equation, which, upon substituting the solution above, reduces to
\begin{equation}
	\frac{1}{3}h''+\frac{1}{r}h'=\frac{\mu}{12\sqrt{2}\pi Gr}+4\lambda h(h^2-v^2).\label{eqPLHiggs}
\end{equation}

\begin{figure}[b]
	\includegraphics{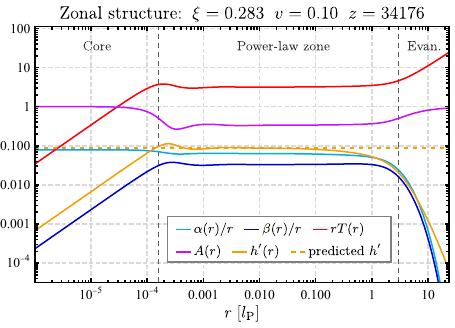}
	\caption{An example of a high-redshift solution, showing the structural separation into three distinct zones (core, power-law, and evanescent). The fields have been rescaled by the appropriate powers of $r$ such that they are approximately constant in the power-law zone. Also included is the predicted power-law zone value for $h'$ calculated from (\ref{eqPL2}). The parameter values for this solution are detailed in Appendix \ref{appData}.}
	\label{figPLZ}
\end{figure}

\begin{figure*}
	\includegraphics{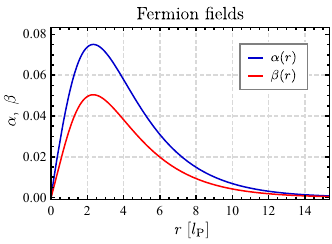}\hspace{7pt}
	\includegraphics{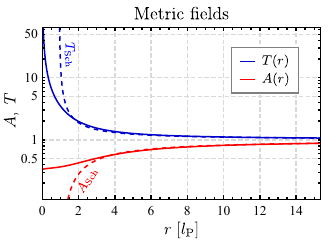}\hspace{7pt}	
	\includegraphics{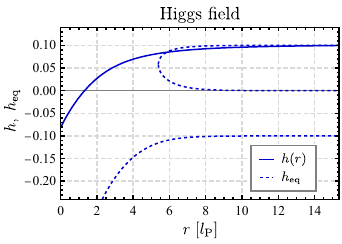}
	\caption{An example of an infinite-redshift state, showing the radial profiles of the fermion, metric and Higgs fields. The power-law zone here extends to $r=0$, with $\alpha\sim r$, $\beta\sim r$, $T\sim 1/r$, $A=1/3$ and $h\sim r+h_0$ at small $r$, as evident. The parameter values for this solution can be found in Appendix \ref{appData}.}
	\label{figPLSoln}
\end{figure*}

To proceed further, we assume that the power-law zone occurs at small $r$ (this should certainly be the case at high redshift), and that $h(r)$ also has a simple power-law dependence. Thus, in order for the Higgs energy density not to contribute to the Einstein equations, we require $h'(r)$ to lead with a power greater than $-1$ at small $r$. This implies that the second term on the right hand side of (\ref{eqPLHiggs}) must be negligible. We can then solve for $h(r)$ to give
\begin{equation}
	h(r)=\frac{\mu}{12\sqrt{2}\pi G}r-\frac{c_1}{2r^2}+c_2,
\end{equation}
where $c_1$ and $c_2$ are constants. From the argument above we are forced to set $c_1=0$, and therefore, in the power law zone, the Higgs field must be approximately linear, i.e.
\begin{equation}
	h(r)=\frac{\mu}{12\sqrt{2}\pi G}r+h_0\label{eqPL2}.
\end{equation}

These expressions can be readily checked by analyzing the structure of high-redshift states, an example of which is shown in Fig.~\ref{figPLZ}. This clearly illustrates the separation of the solution into three distinct zones: the core (in which the fields follow the small-$r$ expansion), the power-law zone (where all fields have approximately power-law dependence on $r$), and the evanescent zone (in which the fermion fields decay exponentially). As predicted, the Higgs field is indeed approximately linear in the power-law zone ($h'=\mathrm{const.}$), and the precise numerical values agree well with those derived in the expressions above. Note that the oscillations within the power-law zone are caused by a fermion self-trapping effect, details of which can be found in ref.\ \cite{Leith2020fermionTrapping}.

As the redshift is increased further, the spatial extent of the core shrinks towards zero, and it is therefore possible to generate infinite-redshift states numerically by replacing the small-$r$ expansion (\ref{eqSmallrStart})--(\ref{eqSmallrEnd}) with the power-law expressions (\ref{eqPL1}) and (\ref{eqPL2}). An example of such an infinite-redshift solution is shown in Fig.~\ref{figPLSoln}. Note that the metric field $T$ diverges at $r=0$, and the state contains a central spacetime singularity. The input values for this solution are $\{\xi,v\}=\{0.28,0.10\}$, and thus we expect the state to lie at the center of the orange spiral shown earlier in Fig.~4. Noting the output parameter values of $m_f=0.468$ and $\omega=0.260$, this is indeed confirmed to be the case.

\section{Alternative parametrizations}\label{appPar}

Recall that, of the three physical parameters $\{\mu,\lambda,v\}$, only two can be freely specified as inputs; the other is fixed by imposing normalization. As previously mentioned, the computationally efficient choice for the two input parameters is $v$ and $\xi=m_H/m_f=2\sqrt{2\lambda}/\mu$. As such, the families of solutions presented in the main text are defined by these values.

\begin{figure}[b]
	\includegraphics{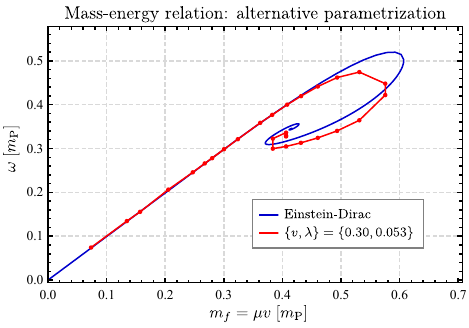}
	\caption{The fermion mass-energy relation for the family of states with parameter values $v=0.30$ and $\lambda=0.053$, along with the corresponding Einstein--Dirac curve for comparison.}
	\label{figConstLambda}
\end{figure}

\begin{center}
	\begin{figure*}[t]
		\includegraphics{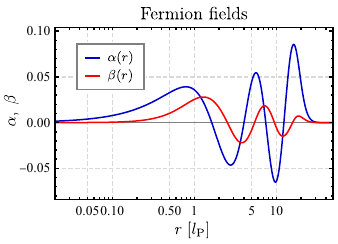}\hspace{7pt}
		\includegraphics{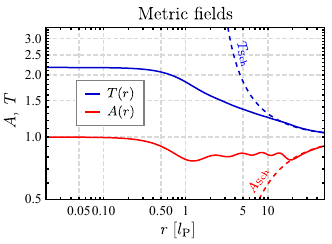}\hspace{7pt}	
		\includegraphics{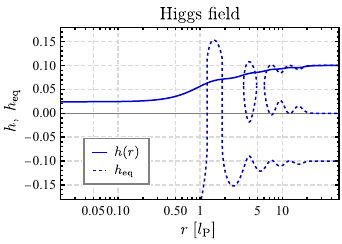}
		\caption{An example of an 8th excited state ($n=8$), showing the radial profiles of the fermion, metric and Higgs fields. There are a total of eight nodes in $\alpha$ and $\beta$ (four in each). The parameter values for this solution can be found in Appendix \ref{appData}.}
		\label{figExcited}
	\end{figure*}
\end{center}

It is important to note that this is not a unique choice, however. It should in principle be possible to choose any two parameters from the set $\{\mu,\lambda,v\}$ (or two independent combinations), with each separate choice defining a distinct family of states. As an example of this, we have generated the family of solutions defined by $\{\lambda=0.053,v=0.3\}$, and the resulting fermion mass-energy curve is shown in Fig.~\ref{figConstLambda}. Note that this contains only a relatively small number of points, since it is significantly more difficult to obtain computationally. Nonetheless, it is clear that this family of states exhibits the expected spiraling behavior, and we find that the curve is indeed parameterized by the central redshift. We expect other parameter choices to produce similar curves. It is important to point out, however, that choosing a new pair of parameters to use does not produce a new set of states; it only reparametrizes the 2-dimensional manifold of solutions.

\section{Excited states}\label{appExcited}

The analysis presented in the main text is limited to ground-state solutions of the Einstein--Dirac--Higgs system. We are also able to obtain excited states, and have found that these are similar in structure to those in the Einstein--Dirac case \cite{FSY1999original, Leith2021excited}. In particular, for each value of the central redshift $z$, there exists a (presumably infinite) tower of excited states, where the $n^{\mathrm{th}}$ excited state contains a total of $n$ nodes in the fields $\alpha$ and $\beta$.

An example of an 8th excited state is shown in Fig.~\ref{figExcited}. Note that the additional oscillations in the fermion fields affect not only the metric but also the Higgs field, since the tilt in the effective Higgs potential is proportional to $\alpha^2-\beta^2$. Note also that the term `excited' implies higher fermion energy only when the system is non-relativistic. For example, it is possible at high redshift for an excited state to have a lower value of $\omega$ than the ground state, and in such cases only the nodal structure can be used to categorize the states.

\end{document}